\documentclass[english,prd,twocolumn,superscriptaddress,showpacs,preprintnumbers,amsmath,amssymb,aps]{revtex4-2}
\usepackage{graphicx}
\usepackage{dcolumn}
\usepackage{bm}

\usepackage{upgreek}
\usepackage{multirow}
\usepackage{lineno}   

\begin{document}


\title{Search for low-mass electron-recoil dark matter using a single-charge sensitive SuperCDMS-HVeV Detector}

\author{M.F.~Albakry} \affiliation{Department of Physics \& Astronomy, University of British Columbia, Vancouver, British Columbia V6T 1Z1, Canada}\affiliation{TRIUMF, Vancouver, British Columbia V6T 2A3, Canada}
\author{I.~Alkhatib} \email{Corresponding author: imran.alkhatib@mail.utoronto.ca} \affiliation{Department of Physics, University of Toronto, Toronto, Ontario M5S 1A7, Canada}
\author{D.~Alonso-González} 
    \affiliation{Departamento de F\'{\i}sica Te\'orica, Universidad Aut\'onoma de Madrid, 28049 Madrid, Spain}
    \affiliation{Instituto de F\'{\i}sica Te\'orica UAM-CSIC, Campus de Cantoblanco, 28049 Madrid, Spain}
\author{J.~Anczarski} \affiliation{SLAC National Accelerator Laboratory/Kavli Institute for Particle Astrophysics and Cosmology, Menlo Park, California 94025, USA}
\author{T.~Aralis} \affiliation{SLAC National Accelerator Laboratory/Kavli Institute for Particle Astrophysics and Cosmology, Menlo Park, California 94025, USA}
\author{T.~Aramaki} \affiliation{Department of Physics, Northeastern University, 360 Huntington Avenue, Boston, Massachusetts 02115, USA}
\author{I.~Ataee~Langroudy} \affiliation{Department of Physics and Astronomy, and the Mitchell Institute for Fundamental Physics and Astronomy, Texas A\&M University, College Station, Texas 77843, USA}
\author{C.~Bathurst} \affiliation{Department of Physics, University of Florida, Gainesville, Florida 32611, USA}
\author{R.~Bhattacharyya} \affiliation{Department of Physics and Astronomy, and the Mitchell Institute for Fundamental Physics and Astronomy, Texas A\&M University, College Station, Texas 77843, USA}
\author{A.J.~Biffl} \affiliation{Department of Physics and Astronomy, and the Mitchell Institute for Fundamental Physics and Astronomy, Texas A\&M University, College Station, Texas 77843, USA}
% ...
\collaboration{SuperCDMS Collaboration}

\date{\today}

\begin{abstract}
We present constraints on low-mass dark matter electron scattering and absorption interactions using a SuperCDMS high-voltage eV-resolution (HVeV) detector. Data were taken underground in the NEXUS facility located at Fermilab with an overburden of 225 meters of water equivalent. The experiment benefits from the minimizing of luminescence from the printed circuit boards in the detector holder used in all previous HVeV studies. A blind analysis of $6.1\,\mathrm{g\cdot days}$ of exposure produces exclusion limits for dark matter-electron scattering cross sections for masses as low as $1\,\mathrm{MeV}/c^2$, as well as on the photon-dark photon mixing parameter and the coupling constant between axionlike particles and electrons for particles with masses $>1.2\,\mathrm{eV}/c^2$ probed via absorption processes.
\end{abstract}

\maketitle


\section{Introduction}

Searches for sub-GeV dark matter (DM) have attracted increasing attention recently as the parameter space of DM above the GeV scale has been strongly constrained by the most sensitive direct detection experiments~\cite{LZ:2024zvo, XENON:2023cxc, DarkSide-50:2022qzh}. In the sub-GeV energy regime, DM candidates include particles produced nonthermally in the early Universe such as MeV-scale DM fermions ($\chi$), as well as keV-scale bosons such as dark photons (DPs) and axionlike particles (ALPs)~\cite{battaglieri2017cosmicvisionsnewideas, EDELWEISS:2013, EDELWEISS:2018}. Cryogenic crystal detectors play an important role in probing this mass range~\cite{supercdmscollaboration2024lightdarkmatterconstraints, arnaud2020first}. SuperCDMS high-voltage eV-resolution (HVeV) crystal bolometers, optimized for $\mathcal{O}(10)\,\mathrm{eV}$ electron recoil energy thresholds and $\mathcal{O}(1)\,\mathrm{eV}$ energy resolution, provide sensitivity to these DM candidates through electron scattering or absorption processes~\cite{Ren_2021}.

Exclusion limits have been produced for the aforementioned models in three previous searches using HVeV detectors~\cite{Agnese_2018,Amaral_2020, supercdmscollaboration2024lightdarkmatterconstraints}. In this study, we present the results of the fourth HVeV DM search (HVeV Run 4), with data taken at the Northwestern EXperimental Underground Site (NEXUS, see Ref.~\cite{Bratrud2025} for more information). Four silicon HVeV detectors (NFC1, NFC2, NFE and NFH), each of mass $0.93\,\mathrm{g}$, were operated inside an upgraded detector housing to eliminate the luminescence events from Printed Circuit Board (PCB) reported in Ref.~\cite{supercdmscollaboration2024lightdarkmatterconstraints}.
Moreover, the NEXUS facility was upgraded with an external lead castle of 10\,cm thickness which allows the suppression of the ambient gamma-ray flux from $^{40}$K and the U- and Th-chains by close to two orders of magnitude~\cite{Bratrud:2024fmj}.
The detectors have varying sensor designs, but the energy resolutions $\sigma_E $ of these detectors are all around $3\,\mathrm{eV}$. The best-performing detector (NFC1) was employed for the DM search, while the remaining detectors served as vetoes to reject multiple-scattered background events. We report exclusion limits on the DM-electron scattering cross section $\sigma_{\mathrm{e}}$~\cite{essig2016directdetectionsubgevdark} down to a mass of $1\,\mathrm{MeV}/c^2$, and limits on the photon-dark photon mixing parameter $\epsilon$ and the ALP-electron coupling constant $g_{\mathrm{ae}}$~\cite{PhysRevD.95.023013} for particle masses as low as $1.2\,\mathrm{eV}/c^2$, based on a net exposure of $6.1\,\mathrm{g\cdot days}$.

\section{Experimental Setup}

The experimental setup is described in Ref.~\cite{ComptonScatterPaper}. Here we only list relevant key features. Four detectors were installed inside a copper holder arranged in two horizontal planes, with two detectors placed side by side on each plane. Only a minimal area of PCB is exposed on the copper holder for wire bonding, minimizing luminescence-induced events. Each detector is made of a high purity silicon crystal, with a square face of $10\,\mathrm{mm}$ side length and $4\,\mathrm{mm}$ thickness. The detector holder sits in a light-tight copper housing which is thermally coupled to the mixing chamber (MC) of a cryogen-free dilution refrigerator. The copper housing has a diameter of $220\,\mathrm{mm}$ and a height of $180\,\mathrm{mm}$, while the MC stage of the fridge was maintained at $11\,\mathrm{mK}$. 

SuperCDMS HVeV detectors use quasiparticle-trap-assisted electrothermal-feedback transition edge sensors (QETs)~\cite{HONG2020163757}, patterned on one side of the silicon crystal, to measure phonons corresponding to energy depositions. The QETs are grouped into two independent readout channels covering the same fractional surface area; an inner channel, which covers a central rectangular region, and an outer channel, which surrounds the inner channel to form a hollow rectangular area. By comparing the signal amplitudes from each channel, one can determine whether the energy deposition of an event occurred closer to the inner or outer region of the detector. 

The QET channels were connected to a SQUID readout circuit and were read out at a sampling frequency of $156.25\,\mathrm{kHz}$. To reduce external magnetic flux coupling into the SQUID circuit, a two-layer magnetic shield configuration was implemented. The inner Amuneal 4K~\cite{amunealA4K} shield encloses the detector area with nearly full coverage, including openings for mechanical and electrical connections, and the outer Metglas (Magnetic Alloy 2705M) blanket~\cite{metglas2705M} covers the full solid angle around the dilution refrigerator, providing additional attenuation of external magnetic fields. A bias voltage of $V_{\mathrm{bias}} = 100\,\mathrm{V}$ was applied across the crystal to induce the Neganov-Trofimov-Luke (NTL) effect \cite{Neganov:1985khw, LUKE1990406}, where phonons are generated through the drifting of electron-hole (eh) pairs liberated by electron recoils or absorptions. The total amount of phonon energy measured by the detector for a single particle interaction, ${E_{\mathrm{ph}}}$, is the sum of the DM-electron recoil energy or absorption energy ${E_{\mathrm{r}}}$ of the interaction and the energy produced from the NTL amplification, as shown in Eq.~(\ref{equ:TotalPhononEnergy}), 
\begin{equation}
    \label{equ:TotalPhononEnergy}
    {E_{\mathrm{ph}}=E_{\mathrm{r}}+n_{\mathrm{eh}} \cdot e \cdot V_{\mathrm{bias}}}
\end{equation}

\noindent where ${n_{\mathrm{eh}}}$ is the number of eh-pairs produced and $e$ is the quantum of electric charge. Equation~(\ref{equ:TotalPhononEnergy}) assumes that the sensitivity to recombination phonons and/or the phonons promptly radiated from the initially released hot charge carriers \cite{ramanathan2020ionization}, is the same as the sensitivity to NTL phonons.

As needed in the experiment, LEDs and a $^{137}\textup{Cs}$ gamma source were used for energy calibration and data selection studies. Four LEDs emitting photons of $630\,\mathrm{nm}$ wavelength (1.97 $\mathrm{eV}$) at room temperature were instrumented inside the copper housing, each shining through a pinhole pointing at the center of one of the detectors, on the side opposite the QETs. The pinholes were covered with an infrared filter (SCHOTT KG3) to block long-wavelength photons beyond the desired LED emission band. These LEDs provided the means for a precise energy calibration. The gamma source $^{137}\textup{Cs}$ was placed outside the refrigerator, at approximately the same height as the detectors and at a radial distance of $65\,\mathrm{cm}$ from them~\cite{ComptonScatterPaper}.

\section{Data Collection and Event Reconstruction}~\label{sec:datacollection}
HVeV Run 4 consists of two separate data collection periods. Period I (February to April 2022) included 12 days of dark matter data collection (10.8 $\mathrm{g\cdot days}$ of raw exposure per detector) with data taken at $100\,\mathrm{V}$ bias and a high-energy calibration period using a $3.14\,\mathrm{MBq}$ $^{137}\textup{Cs}$ gamma source with data taken at both $0\,\mathrm{V}$ and $100\,\mathrm{V}$ bias. Period II focused on calibration with optical photons. Data were taken in a continuous readout mode with the transition edge sensor (TES) current being digitized every $6.4\,\upmu \mathrm{s}$ (described in detail in Ref.~\cite{ComptonScatterPaper}) and stored in the form of 0.5-second long raw readout traces.

A threshold trigger was applied to detect pulses in each readout trace after filtering using a Gaussian derivative kernel. Each detected pulse was further analyzed within a time window of 2048 digitized samples, with the trigger point centered in the window. The main energy estimator is the optimal filter (OF) amplitude of the triggered pulse. The OF requires a pulse template which is constructed from the average of pulses corresponding to events generated by single eh-pairs when the detector is biased at $100\,\mathrm{V}$, and a power spectral density of the noise which is extracted from randomly triggered events after pulse rejection. We use two pulse amplitude estimations: (1) The OF amplitude ($A_{\textup{OF0}}$) where the start time of the pulse in the template is forced to coincide with the trigger time; (2) The maximum OF amplitude ($A_{\textup{OF}}$) when scanning a range within six samples of the trigger time, to account for potential misalignment between the pulse and the OF template. 
$A_{\textup{OF0}}$ is a more suitable estimator for low-energy detector response modeling when the pulse arrival time is known (as in the case of LED data discussed below in Sec.~\ref{calibsec}). It is also essential for an unbiased assessment of the baseline. Both the triggering and amplitude estimations are applied to the summed traces of the inner and outer channels, using equal weighting (1:1) for both channels.. 

Period II of data taking was preceded by the addition of LED sources and filters described in the previous section. A spectrum analyzer was used to measure the energy of the LED photons to be $E_{\mathrm{photon}} = 2.05 \pm 0.02\,\mathrm{eV}$ with the LED immersed in liquid helium. A function generator was used to drive an LED with $O(1 \, \upmu \mathrm{ s} )$ pulses and a repetition rate of 10 Hz. A cross-talk signal between the LED and QET wiring was observed. It was mitigated by minimizing the amplitude swing of the driving voltage through keeping a DC voltage level just below the LED turn-on. Variation of the LED intensity while holding the capacitative cross-talk constant was achieved by holding the signal amplitude constant while the DC offset was changed. An interpolation triggering algorithm was developed to identify triggered LED events and find those that were not triggered because no photons were produced in accordance with Poisson statistics. This LED interpolation trigger exploits the periodicity of the LED pulses by identifying a train of triggered pulses following the exact $0.1\,\mathrm{s}$ spacing in time, interpolating in cases where the threshold trigger did not fire at the time of an LED pulse.

\section{Calibration and Detector Response Modeling}~\label{calibsec}

We calibrated the detector NFC1 with the LED data acquired in Period II. We also extracted parameters of charge trapping (CT) and impact ionization (II) according to the chosen detector response models~\cite{Ponce_2020, Wilson_2024}. Charge trapping occurs when charge carriers are trapped before undergoing the full NTL amplification, resulting in a reduced signal. Impact ionization refers to the process where a drifting charge carrier liberates a previously trapped charge carrier, leading to a larger-than-expected signal. The parameters were extracted using two fits of a single LED dataset: (1) a fit for energy calibration and detector response parameters from the $A_{\textup{OF0}}$ spectrum including the 0 eh-pair peak; followed by (2) a fit for energy calibration parameters from the $A_{\textup{OF}}$ (the energy estimator for the DM search) spectrum, excluding the 0 eh-pair peak (see Fig.~1) while holding the CT fraction $f_{\textbf{CT}}$ and II fraction $f_{\textbf{II}}$ fixed to the values found in the first fit. The motivation behind this two-step fit is that although $A_{\textup{OF}}$ is a more precise energy indicator for most eh-pair peaks, it cannot measure the amplitude of the 0 eh-pair peak correctly under the influence of noise. 

The effects of CT and II are modeled according to Ref.~\cite{Wilson_2024} which provides event rate distributions between adjacent eh-pair peaks for multiple photons hitting the surface of the detector simultaneously (within the timing resolution), each generating one charge carrier pair. This model is defined in the energy domain. It is transformed from the amplitude domain using a quadratic calibration function, with the linear term being dominant and taking into account a second-order correction, while ignoring the higher-order terms assuming their negligible influences.

Due to small changes in the experimental setup to accommodate the LED module installation and possible environmental conditions, the TES bias current ($I_{\textup{TES}}$) was adjusted by less than $7\%$ in Period II compared to Period I, in order to maintain the same TES operating resistance. Since the calibration depends on the TES current, we derive a linear correction to the calibration extracted from Period II measurements before applying it to data from Period I. The correction factor ($\sim0.96$) is derived such that the first eh-pair event amplitudes as measured in non-LED data match between the two periods. Additionally, cross-talk between LED and channel readout wires affects the measured pulse amplitudes, which must be taken into account. To ascertain the amplitude change of the pulse caused by cross-talk, the dependence of the position of the 0 eh-pair peak on LED offset voltage is fit with a linear function and extrapolated to the point where the LED intensity goes to zero (by changing the offset, the cross-talk amplitude stays constant throughout as discussed in Sec.~\ref{sec:datacollection}). This extrapolated peak position is our estimate for the cross-talk contribution, and is subtracted from each OF amplitude before the calibration fits are applied. The 0 eh-pair amplitude itself increases with increasing LED intensity, a feature attributed to the surface trapping effect, where a charge carrier released by a photon near the detector surface promptly recombines instead of traversing the crystal~\cite{HONG2020163757, Wilson_2024}. To account for this surface trapping effect, a shift in energy is introduced in the calibration function that is constant across all eh-pair peaks, but scales with LED intensity.

Three sources of systematic uncertainty on the energy scale were identified: (a) Uncertainty due to cross-talk subtraction; (b) Uncertainty due to the assumption of a quadratic calibration function from amplitude to energy; (c) Uncertainty from the linear gain correction between Period I and II. The first two are taken into account as constant uncertainties; for (a) the estimated cross-talk amplitude uncertainty is converted to energy units, and for (b) we use the maximum residual when comparing the measured eh-pair peak positions to the values predicted by the quadratic calibration function. The systematic uncertainty (c) was calculated separately for each peak, and was quantified by applying a linear gain correction on datasets taken at varying QET biases within $\pm4\,\%$ of the central value in Period II. We take the maximum deviation as the uncertainty estimate for this systematic. Finally, the statistical uncertainties on the linear and quadratic calibration constants are propagated using the calibration function. 

All four uncertainties are added in quadrature, with values summarized for each eh-pair peak in Table~\ref{table:LimitSettingParameters}. The total uncertainty on the phonon energy scale up to and including the fourth eh-pair peak is $\lesssim 1\,\%$. In the following sections, the calibration and detector response parameters and uncertainties from the aforementioned fits are used in analyzing the DM-search data.

\begin{figure}
    \centering

    \includegraphics[width=1.0\linewidth]{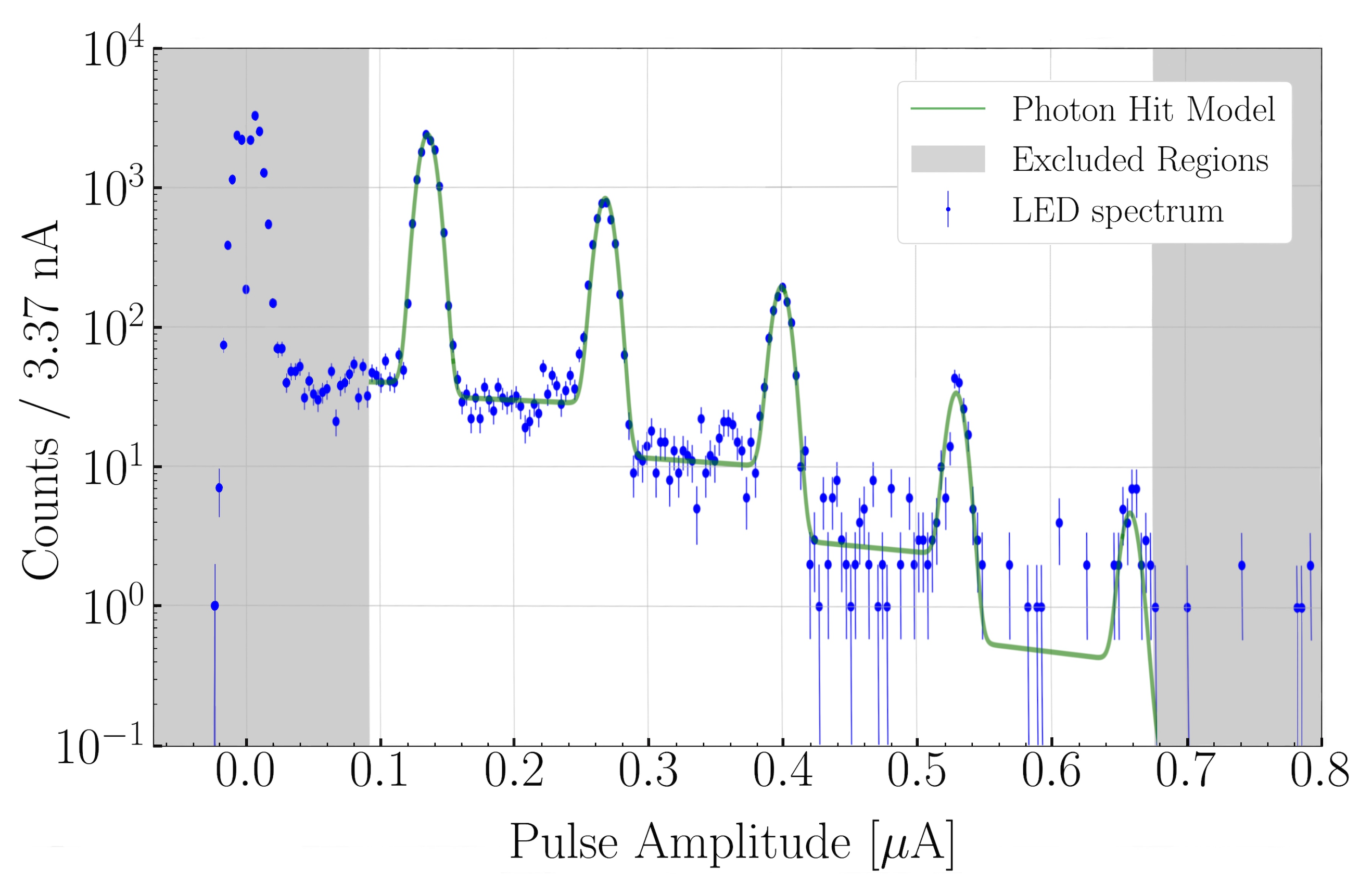}
    \caption{LED calibration data for detector NFC1 and the best-fit photon hit model~\cite{Wilson_2024} for the $A_{\textup{OF}}$ energy estimator. The 0 eh-pair peak is distorted in this  $A_{\textup{OF}}$ amplitude spectrum due to noise fluctuations affecting the OF time offset. The fit is limited to the first five eh-pair peaks because of limited statistics for higher-order peaks.}
    \label{fig:calib}
\end{figure}

\section{Data Selection}

\label{dataselection}

Data selection was performed under a blinding strategy, in which $70\,\%$ of the data remained blinded during the development of the analysis. Specifically, only the first three out of every ten consecutive data files were accessible to the analyzers, with each file corresponding to a 5-minute period. Once the data selection criteria (cuts) were finalized, the same criteria were applied uniformly to the remaining blinded data. The initially unblinded $30\,\%$ data were discarded and not used to produce the final result.

The first stage of quality selections is composed of live-time cuts which exclude periods of data that do not satisfy predefined criteria. The following four live-time selection criteria were developed to ensure only data taken during periods with stable run conditions are used:

\begin{enumerate}
  \item \textbf{Fridge temperature cut}: A cut was applied to exclude data periods when the fridge temperature exceeded $11.1\,\mathrm{mK}$. This cut removes $<1\%$ of the live-time;

  \item \textbf{TES baseline cut}: The average value of the TES current (excluding triggered pulses) was calculated for each readout trace and a $3\,\sigma$ cut was applied to remove outliers. This cut removes $<1\%$ of the live-time;

  \item \textbf{Trigger rate cut}: Elevated trigger rate periods were identified when the trigger rate deviated significantly from the average (i.e., by more than $9\,\sigma$). These periods of time were dominated by time-correlated events incompatible with the random interactions of a stable local halo of DM. A period of ten hours was removed for each observed instance of an elevated trigger rate. This value represents the longest observed elevated rate period in any detector. In all cases, the event rate is consistent with normal periods again after this time. This cut removes approximately $18\%$ of the live-time;
  
  \item \textbf{Coincidence cut}: Events that occur in close temporal proximity are not consistent with the DM signal of interest. Coincidence live-time cuts are therefore defined to exclude any readout trace that contains events within $2^{12}$ time bins (corresponding to $6.5536\,\mathrm{ms}$) of each other (whether they occur in a single detector or across different detectors). This time interval represents half of the pulse analysis window used for event reconstruction. It is chosen to remove specific events that are likely of electronics origin because of their observed characteristic, precise spacing in time. This cut removes $\lesssim1\%$ of the live time.
  
\end{enumerate}
In combination, these live-time selection algorithms remove approximately $19\,\%$ of the raw exposure. 

To ensure the accurate reconstruction of events, a reduced-$\chi^2$ cut is applied. This is based on the reduced-$\chi^2$ value obtained after pulse fitting using the OF described in Sec.~\ref{sec:datacollection}. The amplitude-dependent $3\,\sigma$ upper bound on the reduced-$\chi^2$ is defined on the distribution of reduced-$\chi^2$ and pulse amplitude in the $0\text{ }\mathrm{V}$ $^{137}\mathrm{Cs}$ calibration data, which provides a continuous spectrum of pulse amplitudes. We chose $0\text{ }\mathrm{V}$ $^{137}\mathrm{Cs}$ data over $100\text{ }\mathrm{V}$ background data due to the statistical requirement for enough events at different energies throughout the analysis range. This is justified by the observation of consistent pulse shapes between $0\text{ }\mathrm{V}$ and $100\text{ }\mathrm{V}$ data for pulses with similar amplitudes. The $100\text{ }\mathrm{V}$ LED data of Period II had different noise and cross-talk conditions compared to Period I DM search data, and could not be used for this purpose. 

The signal efficiency of the reduced-$\chi^2$ was estimated from $0\text{ }\mathrm{V}$ $^{137}\mathrm{Cs}$ data, by applying the cut on a subset of the events that are signal-like. This subset of `test' events are deemed signal-like if they pass a pile-up cut that removes time-correlated events, a baseline slope cut removing events with a nonzero slope in the prepulse region and a pulse fall-time cut removing slow fall-time pulses unique to $0\text{ }\mathrm{V}$ datasets which are consistent with local TES saturation \cite{HONG2020163757}. The passage fraction of the $\chi^2$ cut for the remaining events was evaluated in $20\text{ }\mathrm{eV}$ energy bins and taken as an estimate for the cut efficiency. This efficiency is found to be consistent with being energy independent. We take the maximum deviation from the mean as the estimate of the uncertainty of the efficiency, producing a signal survival efficiency $\epsilon_{\chi^2} = 0.95\pm0.02$. 

The energy region of interest (ROI) of this analysis is $85\,\mathrm{eV}$ to $500\,\mathrm{eV}$, which includes the first, second, third and fourth eh-pair peaks (beyond which our limited exposure is not competitive). The lower bound of the ROI is the $5\,\sigma_{\textup{1eh}}$ lower bound of the first eh-pair peak energy, where $\sigma_{\textup{1eh}}$ is the width of the peak. The trigger efficiency in the ROI is estimated from the LED data to be $0.998\pm0.002$, consistent with $100\,\%$. 

\section{Signal Models}

In this analysis, four different DM signals are considered, as described in Refs.~\cite{Essig:2011nj, Hochberg:2016sqx, supercdmscollaboration2024lightdarkmatterconstraints}. There is not a complete understanding of the source of events in the DM-search data, so the observed events are treated under the signal-only hypothesis to derive conservative upper bounds on DM interaction strengths. For DM fermions ($\chi$)-electron scattering with a heavy or light mediator, exclusion limits are set on the $\chi$-electron interaction cross sections $\sigma_{\mathrm{e}}$. For DP and ALP absorptions~\cite{wilson_2023_8370398, wilson_2023_8370457, wilson_2023_8370472}, limits are set on the DP effective mixing parameter ${\epsilon}$ and the axioelectric coupling constant ${g_{\mathrm{ae}}}$, respectively. In each case, it is assumed that the local DM halo with a density of $0.3\,\mathrm{GeV}/\mathrm{cm}^3$ consists exclusively of the candidate particle~\cite{Baxter_2021}. The velocity distribution of the particles is taken from Ref.~\cite{Baxter_2021}, with an average DM velocity of 238\,km/s in the galactic frame, and a galactic escape velocity of 544\,km/s, taking the average speed of the Earth in the galactic frame over a year.

The probability distribution of ${n_{\mathrm{eh}}}$ as a function of ${E_{\mathrm{r}}}$ is determined by an ionization model~\cite{wilson_2023_8370398, wilson_2023_8370498}. In this analysis, the ionization yields were determined using the results from Ref.~\cite{ramanathan2020ionization} and applied in the same way as for the analysis of data from the previous HVeV DM search~\cite{supercdmscollaboration2024lightdarkmatterconstraints}, where the silicon band gap energy used is $1.131\,\mathrm{eV}$~\cite{Stanford:2020xli}.

The CT and II model with parameters $f_{\textbf{CT}}$, $f_{\textbf{II}}$ taken from fit of LED data in Sec.~\ref{calibsec}, is convolved with the NTL phonon energy [${n_{\mathrm{eh}} \cdot e \cdot V_{\mathrm{bias}}}$ term in Eq.~(\ref{equ:TotalPhononEnergy})] to include the effects of charge trapping and impact ionization. Surface trapping which was considered for LED datasets is irrelevant for DM events which would occur throughout the bulk of the substrate. The phonon resolution in the DM search data is taken as the width of the first eh-pair peak (resolution values from LED data are affected by cross-talk). The detector response parameter values used in the DM search are listed in Table~\ref{table:LimitSettingParameters}.

\section{Results}

We follow the same idea as in the previous HVeV DM search~\cite{supercdmscollaboration2024lightdarkmatterconstraints} to take each eh-pair peak region in the energy spectrum as a separate experiment, and decide which peak to use for the final upper limit prior to unblinding the full data files. The upper limits are calculated separately for the first four eh-pair peaks. The peak windows are ${[E_n - 3\sigma, E_n + 3\sigma]}$, where  ${E_n = n \cdot e \cdot V_{\mathrm{bias}} + \langle E_r \rangle}$ is the expected total phonon energy for the $n{\text{th}}$ eh-pair peak given a primary energy deposition with an expectation value $\left< E_r \right>$, and ${\sigma}$ is the median value of the detector resolution estimate. For DPs and ALPs of mass ${m}$, the absorbed energy is taken as ${\langle E_r \rangle =  m\mathrm{c}^2}$. In the case of DM-electron scattering we need to consider that the distribution of recoil energies that may produce $n$ eh pairs (with $n$ ${\leq}$ 4) has a width of less than 18 eV with a $1{\sigma}$ equivalent of less than ${\sim}$5–6 eV \cite{ramanathan2020ionization}, which is considerably less than the peak windows of ${E_n}$ ${\pm}$ ${3\sigma}$ we consider. Therefore, we can use the same approach for DM-electron scattering as for the absorption analyses where we calculate ${\langle E_{r} \rangle}$ as follows. Convolve the DM recoil spectrum with the probability distribution for producing $n$ eh pairs and average the resulting distribution. To avoid biases and look-elsewherelike effects due to separately calculating the upper limit on multiple eh-pair peaks, we preselected the eh-pair peak that is used to compute the final upper limit by selecting the eh-pair peak that produces the strongest constraint for each DM candidate and mass sampled using the $30\,\%$ initially unblinded data. The peak selection choices are applied to the remaining 70\,\% after unblinding to produce the final result.

A likelihood-based limit setting approach~\cite{Algeri_2020} is adopted in this analysis. The unbinned extended likelihood function $L(\mu, \boldsymbol{\theta};\boldsymbol{E})$ for a total of $N$ observed events is constructed as

\begin{equation}
    \label{equ:R4Likelihood}
    {L(\mu, \boldsymbol{\theta}) = \frac{\nu_{\chi}^N e^{-\nu_{\chi}}}{N!} \prod_{i=1}^{N} f_{\chi}(\mu, \boldsymbol{\theta};E_{i}) \prod_{k}\frac{1}{\sqrt{2\pi\sigma_{k}^{2}}} e^{-\frac{(\theta_{k}-\mu_{k})^2}{2\sigma_{k}^{2}}} }
\end{equation}

\begin{equation}
    \label{equ:R4Likelihood2}
    {\nu_{\chi}} = X \int_{a}^b \frac{dR}{dE}(\mu, \boldsymbol{\theta}) \epsilon(E) dE
\end{equation} 

\noindent where ${\mu}$ is the parameter of interest ($\sigma_{\mathrm{e}}$, $\epsilon$, or $g_{\mathrm{ae}}$),  $\nu_{\chi}$ is the expected number of signal events, ${E_{i}}$ is the measured total phonon energy [$E_{\textup{ph}}$ from Eq.~(\ref{equ:TotalPhononEnergy})] of the ${i\mathrm{th}}$ event, and $f_{\chi}$ is the signal probability density function in the $E_{\textup{ph}}$ domain [Eq.~(10) in Ref.~\cite{Wilson_2024}], which also depends on the detector response parameters summarized in Table~\ref{table:LimitSettingParameters} and the ionization model taken from Ref.~\cite{ramanathan2020ionization}. In the Gaussian constraint term, ${\theta_{k}}$ is the ${k{\mathrm{th}}}$ nuisance parameter with an expected mean of ${\mu_{k}}$ and a standard deviation of ${\sigma_{k}}$. In Eq.~(\ref{equ:R4Likelihood2}), ${\epsilon(E)}$ is the DM signal efficiency, $X$ is the exposure after live-time cuts, $a$ and $b$ are the lower and upper boundary of the energy analysis window. Due to the lack of knowledge about either the rate or the energy distribution of the background events, we make the signal-only assumption (taking the expected number of background events ${\nu_{\mathrm{b}}}=0$) and only set conservative upper limits.

Table~\ref{table:LimitSettingParameters} summarizes the nuisance parameters considered in this analysis. Their prior distributions were assumed to be Gaussian, with the mean and standard deviation values taken from the previous sections. The calibration uncertainty is taken into account by introducing a nuisance parameter that simply translates the signal model by an `energy shift', which is constrained by a Gaussian centered at zero (corresponding to the nominal calibration result), with a width equal to the eh-pair-peak-dependent uncertainty estimated at the end of Sec.~\ref{calibsec}.

\begin{table}[ht]
    
    \centering
    \begin{tabular}{|c|c|c|} 
    \hline
     & $\mu_k$ & $\sigma_k$ \\ [0.5ex] 
    \hline
    Detector resolution [eV] & 3.2 & 0.1 \\ 
    \hline
    \multirow{4}{4cm}{\centering Energy shift [eV]} & \multirow{4}{0.5cm}{0.0} & 1 eh: 0.7 \\ 
    &  & 2 eh: 2.0 \\ 
    &  & 3 eh: 3.3 \\ 
    &  & 4 eh: 3.2 \\ 
    \hline
    ${\chi}^2$ cut flat efficiency & 0.95 & 0.02  \\ 
    \hline
    Charge trapping fraction (${\%}$) & 12.3 & 0.5 \\
    \hline
    Impact ionization fraction (${\%}$) & 0.1 & 0.4 \\

    \hline
    
    \end{tabular}
    \caption{Gaussian prior distributions of nuisance parameters. Here $\mu_k$ and $\sigma_k$ represent the mean and standard deviation of the ${k{\mathrm{th}}}$ nuisance parameter, respectively. The calibration uncertainty is taken into account by introducing a nuisance parameter that simply translates the signal model by an `energy shift'.} 
    \label{table:LimitSettingParameters}
\end{table}

\begin{figure}[!ht]
    \centering
    \includegraphics[height=5.6cm]{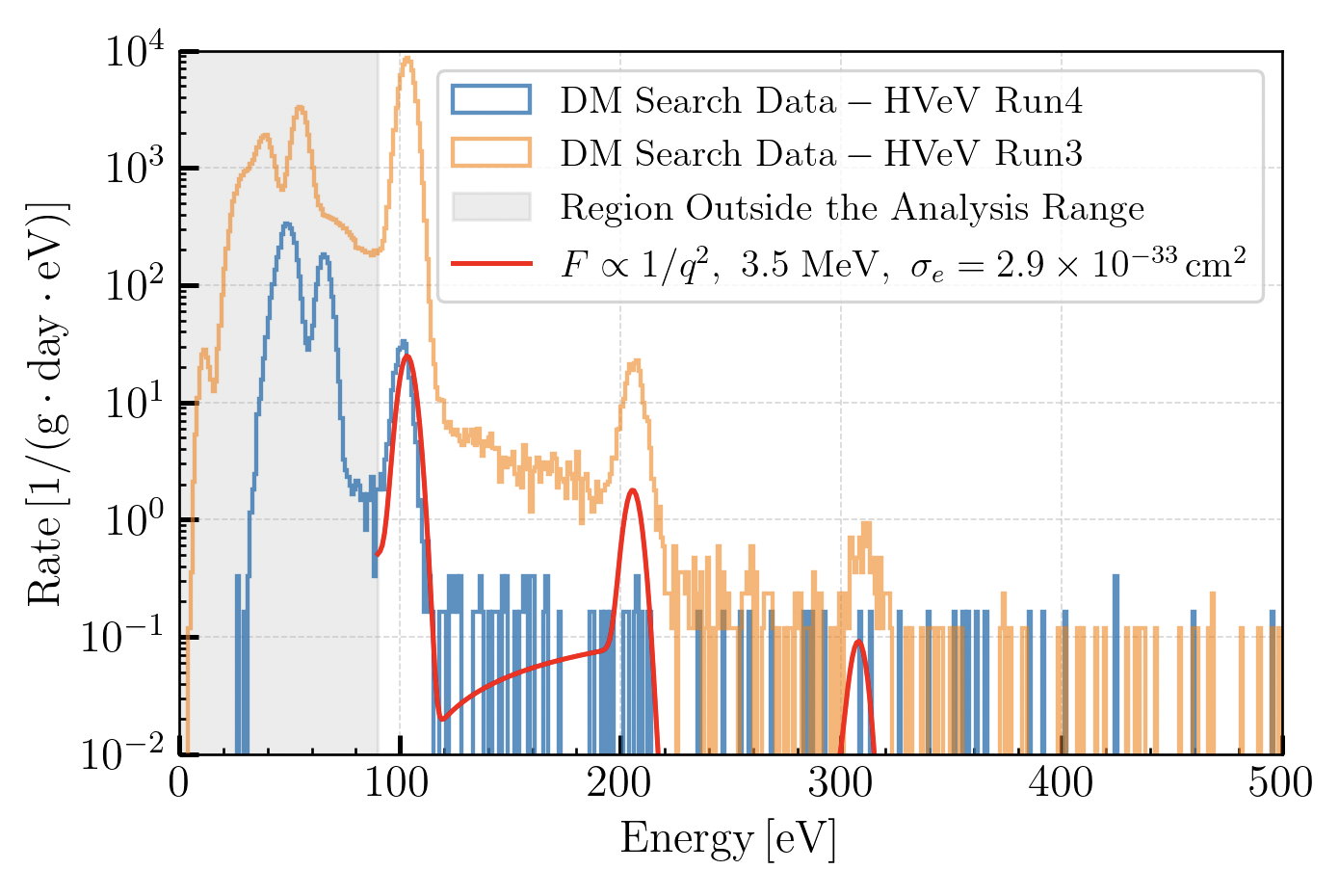}   
    \caption{The blue histogram shows the event rate for the HVeV Run 4 DM-search data after live-time and event-based data selection. The HVeV Run 3 data, after applying all data selection algorithms used in HVeV Run 3, are also shown in orange for comparison. The red curve shows an example of a DM-electron recoil signal model for DM particles with a mass of $3.5\,\mathrm{MeV}/c^2$ interacting through the exchange of a light mediator with a scattering cross section of $2.9 \times 10^{-33}\,\mathrm{cm^2}$. The signal model is shown for the cross section value corresponding to the $90\,\%$ confidence level upper limit produced in this work. The gray-shaded energy range is not considered in this analysis.}
    \label{fig:SpectrumPlot}
\end{figure}

\begin{figure*}[ht]
    \centering
    \begin{minipage}{0.48\textwidth}
        \centering
        \includegraphics[width=\linewidth]{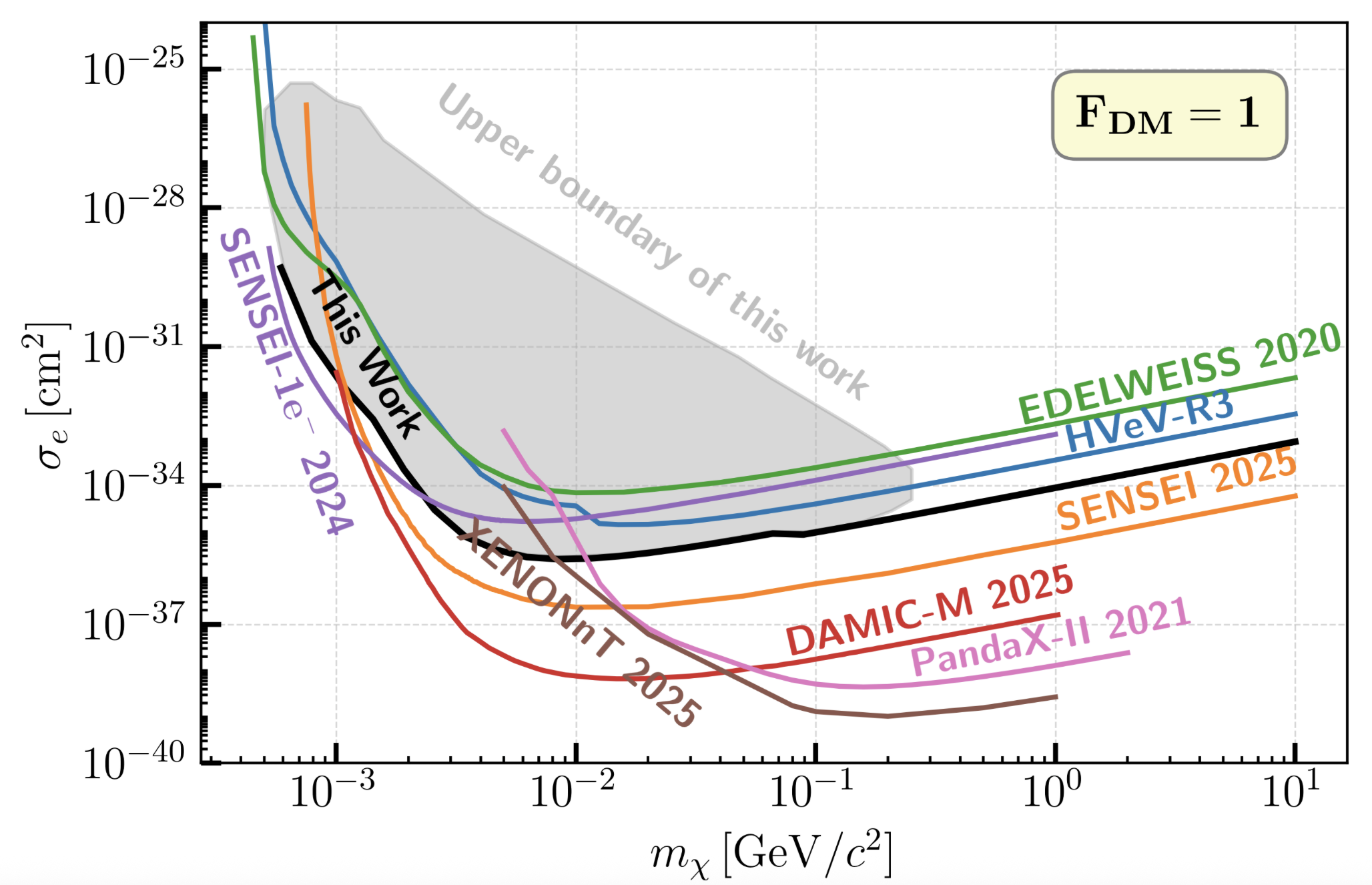} 
    \end{minipage}
    \hfill
    \begin{minipage}{0.48\textwidth}
        \centering
        \includegraphics[width=\linewidth]{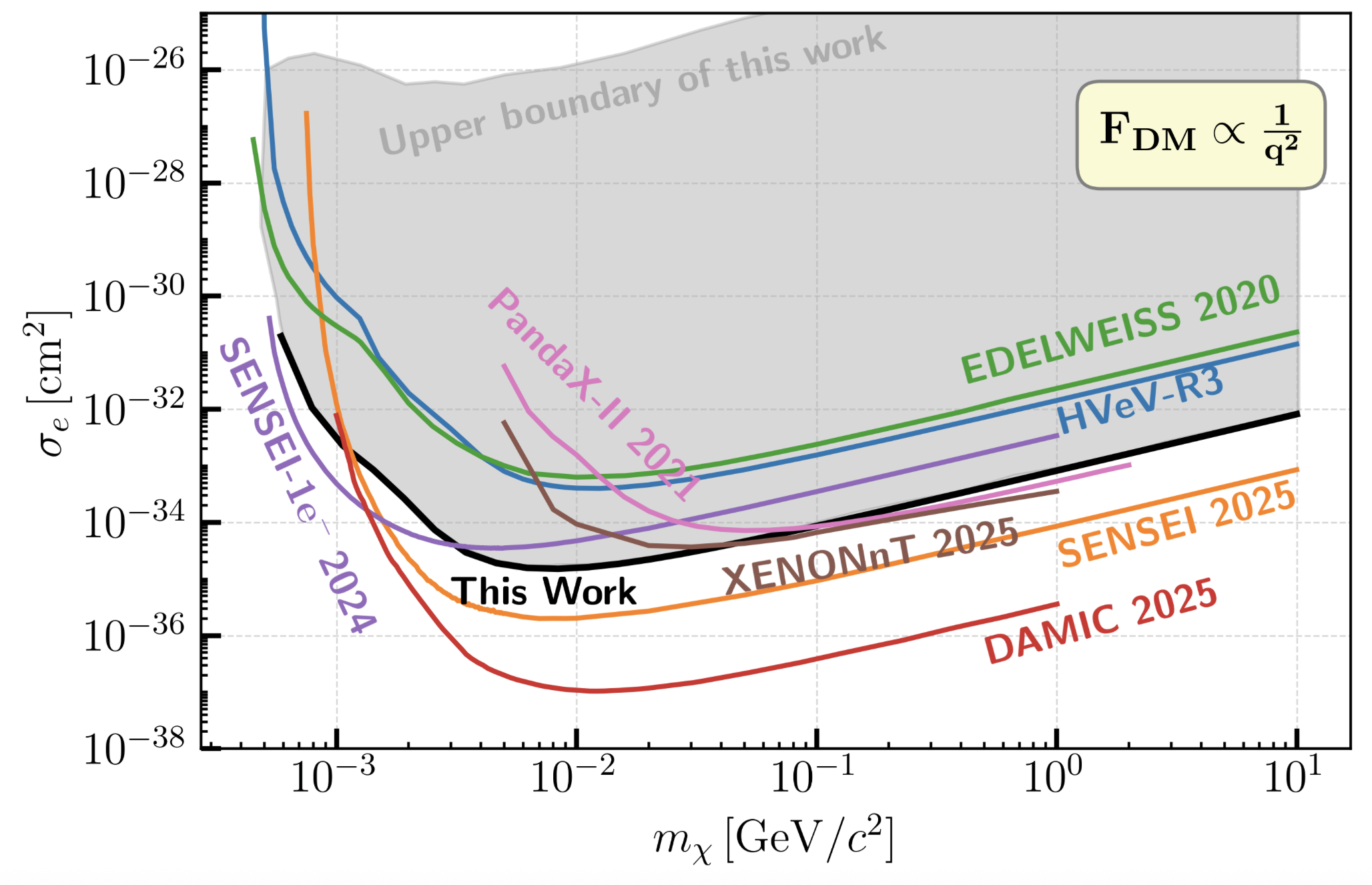} 
    \end{minipage}

    \vspace{0.5cm} 

    \begin{minipage}{0.48\textwidth}
        \centering
        \includegraphics[width=\linewidth]{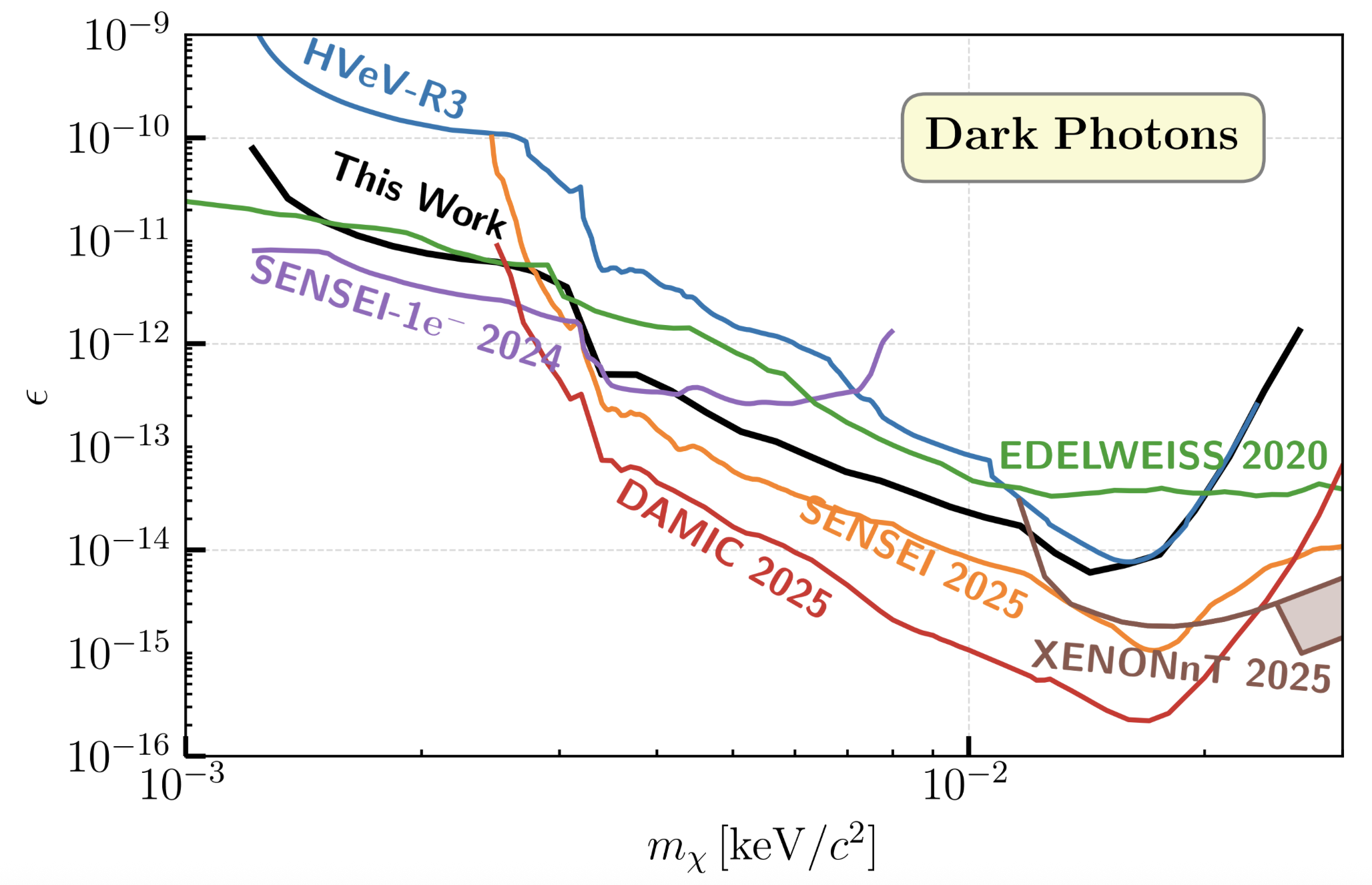}
    \end{minipage}
    \hfill
    \begin{minipage}{0.48\textwidth}
        \centering
        \includegraphics[width=\linewidth]{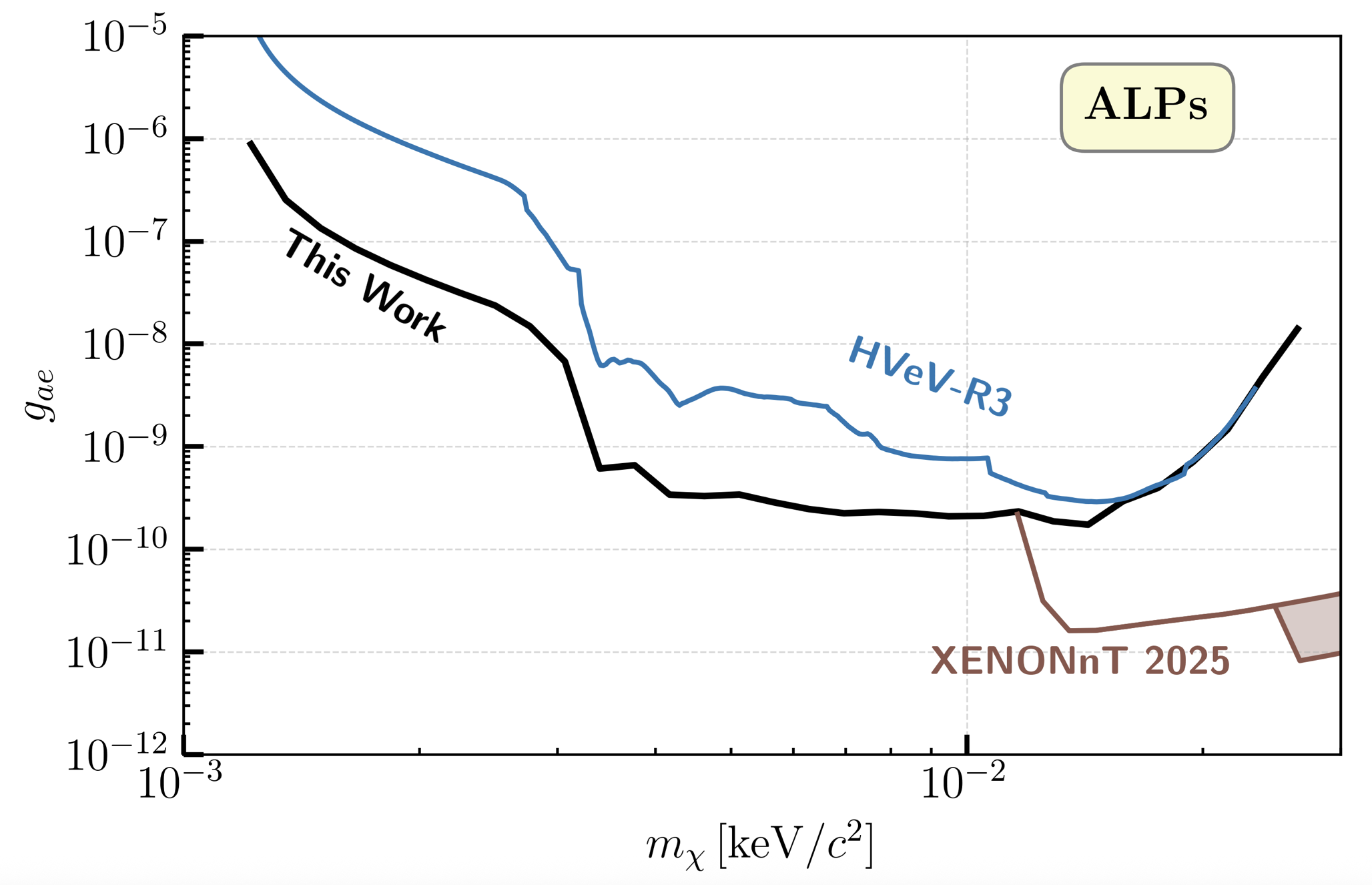}
    \end{minipage}

    \caption{The $90\%$ confidence level exclusion limits from this work (black lines) calculated while ignoring the effects of overburden in comparison with results from other experiments~\cite{supercdmscollaboration2024lightdarkmatterconstraints, SENSEI:2023zdf, arnaud2020first, cheng2021search, XENON:2024znc, DAMIC-M:2025luv} (colored lines) for the $\chi$-electron scattering cross section with DM form factor $F_{\mathrm{DM}} = 1$ (upper left) and  $F_{\mathrm{DM}} \propto 1/q^2$ (upper right), the dark photon kinetic mixing parameter (bottom left) and the axioelectric coupling constant (bottom right). The gray shaded regions in the top plots indicate the estimated exclusion boundaries when considering overburden attenuation \cite{emken2019direct}.}
    \label{fig:FinalLimits}
\end{figure*}

The test statistic ${t(\mu)}$ is defined as

\begin{equation}
    \label{equ:TestStatistic}
    t(\mu) = 
    \begin{cases}
      -2\ln \frac{L(\mu, \boldsymbol{\theta^*})}{L(\hat{\mu}, \boldsymbol{\hat{\theta}})} & \mu \geq \hat{\mu}  \\
      0 & \mu < \hat{\mu}   
    \end{cases}
\end{equation}

\noindent where ${\hat{\mu}}$ and ${\boldsymbol{\hat{\theta}}}$ represent the best fit-values corresponding to the globally maximized likelihood, and ${\boldsymbol{\theta^*}}$ are the best-fit values for a specific values of ${\mu}$. The distribution of ${t(\mu)}$ in the high-statistics limit approaches a ${\chi^2}$ distribution as predicted by Wilk's theorem \cite{cowan2011use}. When the number of observed events is small (${N\leq20}$), we approximate the distribution of ${t(\mu)}$ using Poisson counting results with the nuisance parameters fixed at their mean values, which has been validated by Monte Carlo simulations. 

Due to the attenuation from the overburden, the experiment will not be able to exclude interaction strengths above certain upper boundaries. This Earth shielding effect for $\chi$-electron recoils is estimated following a similar approach as used in the analysis for the previous HVeV DM run~\cite{supercdmscollaboration2024lightdarkmatterconstraints} where a Monte Carlo simulation of the Earth's nuclear stopping power was performed~\cite{emken2019direct}.

After the whole analysis was developed, the remaining $70\,\%$ of the data were unblinded. Figure~\ref{fig:SpectrumPlot} shows the HVeV Run 4 DM-search spectrum after live-time and event-based selections. The 1 eh-pair peak has the dominant number of observed events, with only a few events in the higher order peak regions. In fact, no peak feature is observed after the second eh-pair peak at $\sim200\,\mathrm{eV}$. The red curve in Fig.~\ref{fig:SpectrumPlot} shows one instance of the signal model for DM-electron scattering with a light mediator and DM particle mass of $3.5\,\mathrm{MeV}/c^2$, with the inclusion of the CT and II effects. The cross section of the signal model is chosen to match the $90\,\%$ confidence-level upper limit calculated from the second eh-pair peak of the DM-search spectrum. Figure~\ref{fig:FinalLimits} shows the final constraints on $\sigma_{\mathrm{e}}$, $\epsilon$, and $g_{\mathrm{ae}}$ calculated from the DM search data shown in Fig.~\ref{fig:SpectrumPlot} at $90\%$ confidence level. The black solid lines show the upper limits calculated while ignoring the effects of overburden. The gray shaded regions in the top plots indicate the exclusion boundaries estimated from the attenuation by the overburden. For the dark absorption limits the effect of the Earth attenuation becomes relevant only for very high values of $\epsilon$ and $g_{\mathrm{ae}}$ that are outside the plotting range. Limits from other direct DM search experiments~\cite{supercdmscollaboration2024lightdarkmatterconstraints, SENSEI:2023zdf, arnaud2020first, cheng2021search, XENON:2024znc, DAMIC-M:2025luv} are also shown in colored lines for comparison. 

\section{Discussion and Outlook}

The background event rate in HVeV Run 4 has been reduced by at least two orders of magnitude compared with the previous HVeV DM search~\cite{supercdmscollaboration2024lightdarkmatterconstraints} in the ROI, due to the replacement of the detector holder that removed most of the PCB material near the detectors. Consequently, upper bounds have improved significantly in Run 4 compared to Run 3.

Taking advantage of the lower background event rate during this experiment, for the first time we observed elevated trigger rate periods as described in Sec.~\ref{sec:datacollection}, and developed a live-time cut to remove them. The nature of these high-rate events is still under investigation.

This analysis is limited by the lack of a background model, due to the fact that the background sources and rates remain poorly understood. One major hypothesized source of background events are charge leakage events. When a high voltage is applied across the detector, individual charge carriers could tunnel from the electrode into the crystal bulk~\cite{osti_1462788}, generating events within a similar energy range as single eh-pairs. 

Many cryogenic low-threshold experiments observe sharply rising event rates of yet unknown origins below a few hundred eV \cite{adari2022excess}. This background, usually referred to as low-energy excess (LEE) events, has been reported to be dominated by nonionizing events~\cite{EDELWEISS:2016nzl}. Applying HV pushes our signal (consisting of ionizing events) to higher energies through the NTL effect while not affecting the LEE. This minimizes the LEE background contribution in our energy region of interest.

In summary, HVeV Run 4 confirmed the removal of one major external background source in previous HVeV runs, and produced improved exclusion limits with a likelihood-based limit setting method. A main goal for the future experiments is to investigate and model the background components, and further reduce them. The following HVeV run has already taken place in 2024 at a cryogenic underground test facility hosted at the SNOLAB, the analysis of which is an ongoing effort.

\section*{ACKNOWLEDGMENTS}

Funding and support were received from the National Science Foundation, the U.S. Department of Energy (DOE), Fermilab URA Visiting Scholar Grant No. 15-S-33, NSERC Canada, the Canada First Excellence Research Fund, the Arthur B. McDonald Institute (Canada),  the Department of Atomic Energy Government of India (DAE),  J. C. Bose Fellowship grant of the Anusandhan National Research Foundation (ANRF, India) and the DFG (Germany) - Project No. 420484612 and under Germany’s Excellence Strategy - EXC 2121 ``Quantum Universe" – 390833306 and the Marie-Curie program - Contract No. 101104484. Fermilab is operated by Fermi Forward Discovery Group, LLC,  SLAC is operated by Stanford University, and PNNL is operated by the Battelle Memorial Institute for the U.S. Department of Energy under Contracts No. DE-AC02-37407CH11359, No. DE-AC02-76SF00515, and No. DE-AC05-76RL01830, respectively. This research was enabled in part by support provided by Compute Ontario (computeontario.ca) and the Digital Research Alliance of Canada (alliancecan.ca).

\section*{DATA AVAILABILITY}

The data that support the findings of this article are not publicly available. The data are available from the authors
upon reasonable request.

\bibliography{apssamp}

\end{document}